\documentstyle[aps,prl,multicol,epsf,amssymb]{revtex}
\input epsf

\def\dpl{\epsilon^{pl}}
\def\ddpl{\dot\dpl}

\def\del{\epsilon^{el}}

\begin{document}
\bibliographystyle{prsty}

\tightenlines
\draft

\title {
Rearrangements and Dilatancy for Sheared Dense Materials
}

\author{Ana\"el Lema\^{\i}tre$^{(1,2)}$}
\address{
$^{(1)}$Department of physics, University of California, Santa Barbara, California 93106, U.S.A.\\
$^{(2)}$CEA --- Service de Physique de l'\'Etat Condens\'e,
Centre d'\'Etudes de Saclay, 91191 Gif-sur-Yvette, France
}
\date{\today}
\maketitle
\begin{abstract}
Constitutive equations are proposed for dense materials, 
based on the identification of two types of free-volume activated 
rearrangements associated to shear and compaction.
Two situations are studied: the case of an amorphous solid in a stress-strain test,
and the case of a lubricant in tribology test.
Varying parameters, strain softening, shear thinning,
and stick-slip motion can be observed.
\end{abstract}

\begin{multicols}{2}

From food to beauty products, from our bone joints to the gouge
of tectonic faults, the dynamical properties of dense materials determine
important and ubiquitous phenomena that govern our life.
The idea emerged recently, that some sort of universality might be at work in
structural systems, ranging from glasses to granular materials.~\cite{ohern01}
In this letter, I hypothesize that universality results from
the crucial role played by structural rearrangements.
The approach I propose relies on the so-called 
shear transformation zone (STZ) theory, 
introduced to account for elasto-plastic 
transitions in amorphous solids.~\cite{falk}
STZ theory provides a scheme for rearrangement kinetics
in a dense material: Macroscopic deformation is seen as result of
local free-volume activated rearrangements at a mesoscopic scale.
Local states are introduced, related to the local orientations of contact network.
Mean-field equations of motion for those local states 
lead to macroscopic constitutive equations.

STZ theory is originally designed for solids,~\cite{falk}
but it has been shown recently that it could be adapted to the case
of granular materials and successfully account
for important features of granular flows.~\cite{lemaitre01a}
It is shown here that complementing STZ theory with free-volume kinetics
permits to account for important features of both amorphous solids
(strain softening) and dense liquids (stick-slip) within the same framework.
In previous works on STZ theory free-volume appears as a parameter,
although it is obviously a state variable 
that varies as the system dilates or contracts.
Free-volume kinetics is inspired by recent results on granular materials
submitted to vertical tapping: local rearrangements
associated to a buckling instability of chain forces have been shown to 
account for logarithmic relaxation of free-volume.~\cite{vfgm}
The micro-structural picture proposed for dense materials
involves two types of rearrangements,
associated respectively to shear motion and to free-volume relaxation.
The interplay between those two mesoscopic processes and associated timescales
is shown to account for important properties of sheared materials.

To emphasize the analogy between solids and dense liquids,
two important experimental tests are considered:
the case of a stress-strain test performed on a plastic material~\cite{hassan95}
and the case of an (overdamped)
tribology test 
where a dense liquid or a granular material is sheared.~\cite{ssglass,ssglassc,ssgm}
In the former case, the resulting constitutive equations are shown 
to account for strain softening. In the latter, they account for 
a transition between steady sliding and stick-slip motion.
The picture is completed with the case of constant stress experiments (creep tests).

Let me now introduce a basic mechanism for rearrangement kinetics,
along the lines of STZ theory.~\cite{falk}
The present approach remains at a mean-field level.
The overall shear deformation is denoted $\epsilon$
and decomposed into an elastic and a plastic part: $\epsilon=\del+\dpl$.
The elastic deformation is proportional to the shear stress $\sigma=\mu\del$;
the plastic deformation results from local, irreversible transformations of the contact network.
A local rearrangement involves several molecules at a mesoscopic level.
A shear transformation zone is defined as a locus within the material where
an elementary shear is made possible by the local conformation
of neighboring molecules.
An essential remark that lies at the root of STZ theory is that
once a rearrangement has occurred,
some contacts break, some others are formed, and
the molecules involved cannot shear further in the same ``direction'',
although they might shear backward.
This leads to identifying pairs of types of arrangements, that are transformed 
into one another by a local, elementary shear.
To simplify the theory, one pair of orientations
is considered, corresponding to the principal axes of the stress tensor.
An elementary transformation can be sketched as follows,
\begin{center}
\begin{picture}(100,30)(0,0)
\put(50,30){\makebox(0,0){\large$R_+$}}
\put(50,0){\makebox(0,0){\large$R_-$}}
\put(0,0){\makebox(100,30){
\epsfxsize=100\unitlength
\epsffile{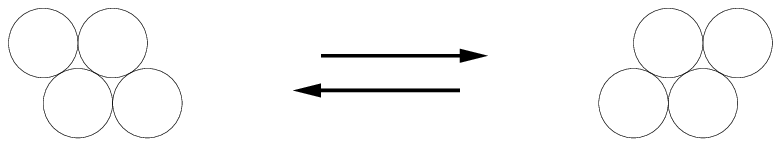}
}}
\end{picture}
\end{center}
\label{fig:stz}
although an actual rearrangement involves in general more than four grains.
Populations (number densities) of arrangements 
of each type are denoted $n_\pm$, and $R_\pm$ denote the rates
at which a $\pm\to\mp$ transformation occurs.
The rates $R_\pm$ depend on force and free-volume fluctuations and will be defined further.
The macroscopic plastic shear is evaluated from the balance between both types 
of elementary motion:
\begin{equation}
\label{eqn:stz:ddpl}
\ddpl = {\cal A}_0\,(R_+n_+-R_-n_-)
\quad,
\end{equation}
where some constant ${\cal A}_0$ has been introduced.
Equation~(\ref{eqn:stz:ddpl}) must be complemented by equations of motion 
for the populations $n_\pm$. They are written of the form,~\cite{falk}
\begin{equation}
\label{eqn:stz:npm}
\dot n_\pm = R_\mp n_\mp - R_\pm n_\pm +\sigma\,\ddpl\,({\cal A}_{c} - {\cal A}_{a}\,n_\pm)
\quad.
\end{equation}
${\cal A}_{c}$ and ${\cal A}_{a}$ are constants.
The first two terms on the rhs account for the internal reconformations of STZ's,
while the last term introduces a coupling with the mean flow:
At the macroscopic scale, the flow constantly stirs the molecules,
thus creating and destroying local configurations.
The rate at which the macroscopic flow induces new configurations
is estimated as the ratio of the overall work, $\sigma\,\ddpl$,
over by some (constant) typical normal force 
(the force required to modify a contact).


Transformation rates $R_\pm$ depend on volume and force fluctuations.
A local reconformation occurs if some volume is available around a transformation zone,
{\it and} if the local force network is appropriately oriented.
In this work, I assume that the probabilities associated with 
those two types of fluctuations factorize.
This differs from the prescription given in~\cite{falk}.
An exponential distribution of volume fluctuations is assumed,
and $v_0$ denotes the typical free-volume per grain needed for a rearrangement.
The probability that sufficient free-volume is localized around a given STZ is
proportional to $\exp\left({-{v_0/v_f}}\right)$,
where $v_f$ is the average free-volume.
The distribution of forces is also expected to be exponential.
Shear stress $\sigma$ introduces a bias of the force network.
The probability that some strong local bias of the force network is realized 
and promotes shearing in the $\pm\to\mp$ direction is
estimated to be proportional to $\exp\left({\pm{\sigma/\bar\mu}}\right)$,
where $\bar\mu$ measures some typical force of the stress network.
In the following, $\bar\mu=1$ is imposed, which fixes the unit of forces.
The rates are thus written,
$$
R_\pm = R_0\,\exp\left({-{v_0/v_f}}\right)\,\exp\left({\pm{\sigma}}\right)
\quad,
$$
with $R_0$, the update frequency of microscopic processes.
This choice of an expression for the rates $R_\pm$ will be discussed further,
but other choices are possible that lead to similar results.~\cite{note1}

At this point, free-volume appears as a parameter.
However, this picture is now complemented with free-volume kinetics.
Free-volume relaxation results from tiny reconformations of molecules into 
better local packings.~\cite{vfgm}
A second type of rearrangement is thus introduced,
which is essentially isotropic
--- hence does not require any particular orientational zone to occur ---
but which is otherwise treated in a similar and consistent way.
The typical volume fluctuation needed for a local collapse is denoted $v_1$;
free-volume relaxation occurs
at a rate proportional to $\exp(-v_1/v_f)$.
Note that $v_1$ is expected to be smaller than $v_0$
as an improvement of the local packing
should require reconformations of a smaller scale than shear motion.
In the absence of shear ($\sigma=0$), and at $0$ temperature,
the equation of motion for free-volume is expected to reduce to, 
$
\dot v_f = -E_1 \exp(-v_1/v_f)
$,
which accounts for a slow (logarithmic) relaxation of $v_f$.~\cite{vfgm}
When the material shears, some dilatancy is expected.
This is due to the fact that macroscopic shear motion brings 
about new random configurations of molecules.
In new arrangements, the local free-volume is not optimized.
This feature must be incorporate in free-volume dynamics {\it via} a creation term.
To estimate this creation term, the following argument is proposed:
the average dilatancy due to shear deformation
is viewed as a transfer of a fraction
of the energy $\sigma\delta\epsilon^{pl}$ into $P\delta v_f$
by some sort of levering effect within the contact network.
Pressure dependence is incorporated in constants.
When the system deforms by $\delta\epsilon^{pl}$,
the average variation $\delta v_f$ of the free-volume
is thus expected to be proportional to $\sigma\delta\epsilon^{pl}$.
Equation of motion for $v_f$ results from the interplay between local 
relaxation and shear-induced dilatancy. It is written:
\begin{equation}
\label{eqn:vf}
\dot v_f = -E_1 \exp\left[-{v_1/v_f}\right]+{\cal A}_v \sigma\ddpl
\quad.
\end{equation}
In the following, constant ${\cal A}_v=1$ is imposed,
which fixes the unit of $v_f$.

Before proceeding further, equations~(\ref{eqn:stz:ddpl}) and~(\ref{eqn:stz:npm})
are written in a more appealing form. Following~\cite{falk}, variables
$$
\Delta = {n_--n_+\over n_\infty}\quad,\quad {\rm and,}\quad
\Lambda = {n_++n_-\over n_\infty}
$$
are introduced,
along with the rescaled parameters $n_\infty = {2 {\cal A}_c / {\cal A}_a}$,
$\epsilon_0 = {{\cal A}_0\,{\cal A}_c/{\cal A}_a}$,
$\gamma = {\cal A}_0\,{\cal A}_c$, and $E_0 = 2\epsilon_0\,R_0$.
It comes,
\begin{eqnarray}
\label{eqn:stzdil:1}
\ddpl&=&E_0\,\exp\left[{-{v_0/v_f}}\right]\,
\left(\Lambda\,\sinh(\sigma)-\Delta\,\cosh(\sigma)\right)\\
\label{eqn:stzdil:2}
\dot\Delta&=&{\ddpl\over\epsilon_0}\,\left(1- \gamma\,\sigma\,\Delta\right)\\
\label{eqn:stzdil:3}
\dot\Lambda&=&\gamma\,\sigma\,{\ddpl\over\epsilon_0}\,\left(1-\Lambda\right)
\end{eqnarray}

Variables $\Lambda$ and $\Delta$ represent respectively the total normalized density
of STZ's and the bias between the populations $n_\pm$.
In the absence of shear motion, the dynamics of $\Lambda$ and $\Delta$ freeze,
while $v_f$ undergoes some ``autonomous'' relaxation:
those two sets of internal variables are associated to very different types of memory. The initial values given to those variables account for the way the 
system has been prepared. 

Let me comment on initial values used for free-volume.
In the absence of shear, $\sigma=0$, the system still undergoes 
free-volume relaxation.
If the picture of a dense material was complete, a reasonable initial state
should correspond to values of $v_f$ for which the factor
$\exp(-v_1/v_f)$ (hence $\exp(-v_0/v_f)$) is vanishingly small:
the system is an aged glass.
However, for a finite temperature,
free-volume is expected to equilibrate at some non-vanishing value.
Temperature dependence is not included in the current approach.
In this letter, to circumvent this question,
the initial condition for $v_f$ is a fraction of $v_0$,
which is small, but not as small
as would be expected from an aged glassy material.
In fact, for very small initial values of the free-volume, a large stress
is necessary to initiate deformation. This produces behavior analogous
to the so-called memory peak. Those aspect are not considered here.

I first describe briefly the resulting behavior in a constant stress experiment 
(creep test), 
before discussing the case of constant strain-rate and tribology experiment.

\begin{figure}
\narrowtext
\begin{center}
\unitlength = 0.0022\textwidth
\begin{picture}(200,100)(-5,0)
\put(0,0){\makebox(200,100){\epsfxsize=220\unitlength\epsffile{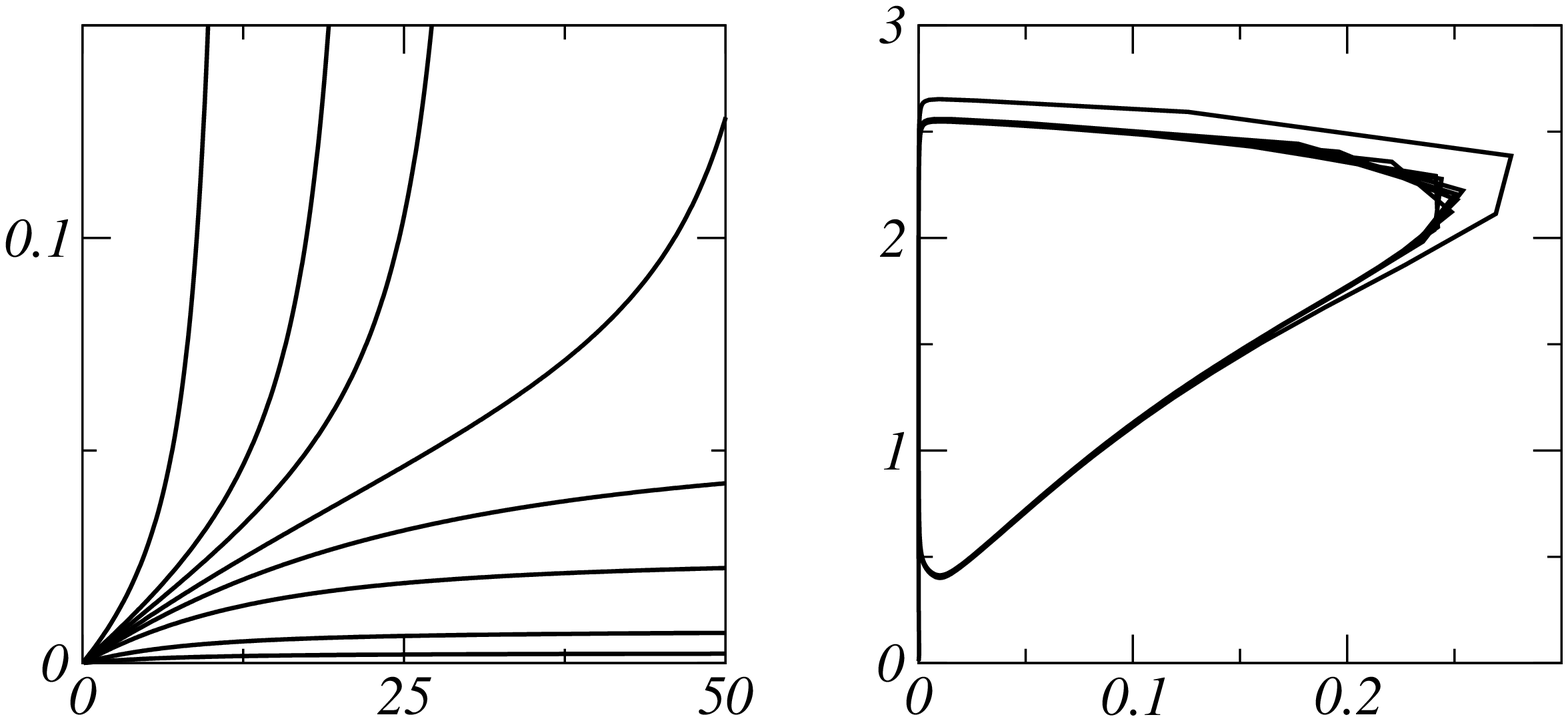}}}
\put(-5,85){\makebox(0,0){$\epsilon$}}
\put(80,0){\makebox(0,0){$t$}}
\put(105,85){\makebox(0,0){$\sigma$}}
\put(190,0){\makebox(0,0){$\ddpl$}}
\end{picture}
\end{center}
\caption{
Left: Creep test, deformation as a function of time at fixed $\sigma$.
Parameters $E_0=E_1=0.1$, $v_0=0.5$, $v_1=0.4$,
$\gamma=10$, $\epsilon_0=1$, and initial conditions $\Delta=0$, $\Lambda=0.01$,
and $v_f=0.25$. From bottom to top, $\sigma=2$, 3, 3.7, 3.9, 4.1, 4.2, 4.5.
Right: Friction force ($\propto\sigma$) as a function of the velocity 
of the upper plate ($\propto\ddpl$) during stick-slip motion
as shown figure~\protect\ref{fig:ssl} (right).
}
\label{fig:2}
\end{figure}

Stress $\sigma$ is fixed. 
To simplify the analysis, I separate equation~(\ref{eqn:stzdil:1}-\ref{eqn:stzdil:3})
from equation~(\ref{eqn:vf}).
Equations~(\ref{eqn:stzdil:1}-\ref{eqn:stzdil:3}) admit two types of solutions:
multiple jammed states ($\ddpl=0$), provided $\Delta=\tanh(\sigma)\Lambda$,
and a flowing regime ($\ddpl\ne0$), with $\Lambda=1$ and $\Delta=1/\gamma\sigma$,
whence
$$
\ddpl= E_0\exp[-v_0/v_f](\sinh(\sigma)-\cosh(\sigma)/(\gamma\sigma)).
$$
The jammed states are stable for small values of the applied stress,
and the two types of solutions exchange stability when the yield stress
$\sigma_y$ is reached, with $\sigma_y$ solution of
$\sigma\tanh(\sigma)=1/\gamma$.~\cite{falk}
The rhs of equation~(\ref{eqn:vf}) is always negative if
$\sigma<\sigma_c$, where $\sigma_c$ is the solution of
$E_0\,(\sigma\,\sinh(\sigma)-\cosh(\sigma)/\gamma)=E_1$.
In this case, the system remains ``solid'', free-volume relaxes,
hence $\ddpl$: the material creeps.
Above $\sigma_c$, equation~(\ref{eqn:vf}) admits one unstable fixed point:
depending on its initial value, $v_f$ either decays or diverges.
In this latter case, the weights $\exp(-v_1/v_f)$ and $\exp(-v_0/v_f)$
saturate to $1$, and $\ddpl$ goes to a constant.
The latest stages of the dynamics, when $v_f$ gets large, are beyond the scope
of the current work.
Various creep curves are shown figure~\ref{fig:2}-(left) which
compare with experiments presented in~\cite{hassan95}.

In a stress-strain test, $\dot\epsilon$ is fixed.
The shear stress is determined by $\sigma=\mu\del$,
or $\dot\sigma=\mu(\dot\epsilon-\ddpl)$.
Let me show that the same relation appears for overdamped stick-slip.
A layer of liquid is sheared between two plates.
The gap between the plates is denoted $a$, and the surface of contact $S$.
A pulling force is exerted on the upper plate, with a spring of stiffness $k$,
pulled at velocity $V$.
If the inertia of the upper plate is neglected: $F=k(Vt-x)$.
The position $x$ of the upper plate is related 
to the deformation of the liquid by: $\dot x=2 a \ddpl$, 
and the shear stress is $\sigma=F/S$, whence, $\dot\sigma=(k/S)(V-2a\ddpl)$, 
which is of the form $\dot\sigma=\mu(\dot\epsilon-\ddpl)$.
Identical equations govern both systems, although, remarkably, 
a solid is usually considered in the former case, and a liquid in the latter.

In the following, $\Lambda=1$ is taken as initial condition.
This is consistent with the fact that stress-strain curves (or stiction peaks)
present well-defined features after preparation of the sample.
Since $\Lambda$ saturate to $1$ after some shearing, it cannot govern some repeatable
test, but accounts for the evolution of the material during preparation.
\begin{figure}
\narrowtext
\begin{center}
\unitlength = 0.0022\textwidth
\begin{picture}(200,200)(-5,0)
\put(0,0){\makebox(200,200){\epsfxsize=220\unitlength\epsffile{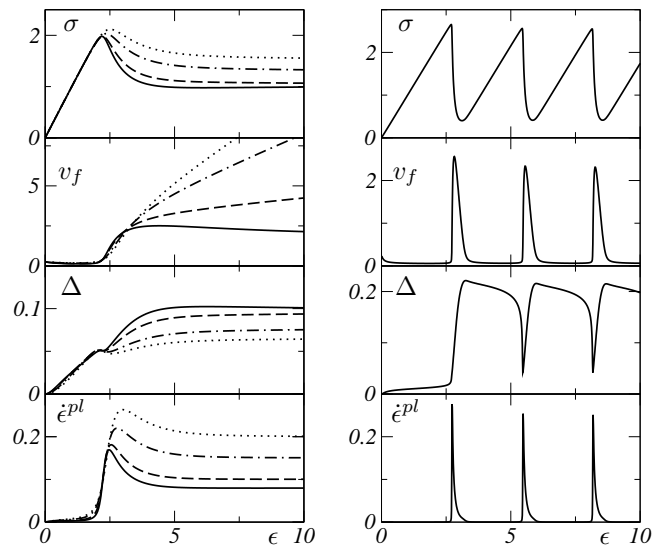}}}
\put(12,183){\makebox(0,0){$\sigma$}}
\put(12,132){\makebox(0,0){$v_f$}}
\put(12,95){\makebox(0,0){$\Delta$}}
\put(12,53){\makebox(0,0){$\ddpl$}}
\put(125,183){\makebox(0,0){$\sigma$}}
\put(125,132){\makebox(0,0){$v_f$}}
\put(125,95){\makebox(0,0){$\Delta$}}
\put(125,53){\makebox(0,0){$\ddpl$}}
\put(80,10){\makebox(0,0){$\epsilon$}}
\put(192,10){\makebox(0,0){$\epsilon$}}
\end{picture}
\end{center}
\caption{
Behavior of sheared materials for fixed strain rate $\dot\epsilon$.
Parameters of the theory are, $E_0=0.1$, $E_1=0.1$, $v_0=0.5$, $v_1=0.4$,
$\gamma=10$, $\epsilon_0=1$. Initial conditions, are $\Delta=0$, $\Lambda=1$,
$v_f=0.25$.
Left: strain softening, for $\dot\epsilon=0.08$ (solid line),
$0.1$ (dashed line), $0.15$ (dot-dashed line), $0.2$ (dotted line).
(Those values are defined up to a time constant, and do not need 
to compare quantitatively with actual values.)
Divergence of $v_f$ is seen for the three largest values of $\dot\epsilon$.
Right: stick-slip motion obtained for $\dot\epsilon=0.01$.
}
\label{fig:ssl}
\end{figure}

Example of behavior resulting from equations~(\ref{eqn:vf}-\ref{eqn:stzdil:2})
is shown figure~\ref{fig:ssl}.
For the large values of $\dot\epsilon$,
the system follows a stress-strain curve which present strain-softening.
(Note that this is for the chosen of parameters, 
and that strain hardening can be seen for other values.)
In the language of tribology, a stiction peak leads to steady sliding.
Softening can be understood as follows:
During a first stage of the dynamics, 
plastic deformation $\ddpl$ remains zero, or very small,
either because, $\sigma$ is below the yield stress, or below the free-volume
remains small. The shear force increasing, transient creeping
motion induces dilatancy,
thus triggering plastic deformation, further increase of $v_f$,
and $\sigma$ starts to decrease with the increasing activation
factor $\exp(-v_0/v_f)$.
For the largest values of $\dot\epsilon$ (here, $\dot\epsilon=0.1, 0.15, 0.2$),
the free-volume diverges, leading the system to melting or break-up.
For intermediate values (here, $\dot\epsilon=0.08$) the system converges
to steady sliding with finite free-volume.
For small $\dot\epsilon$, stick-slip motion is observed,
sliding peaks being associated to sudden rises of the free-volume.
The friction force is multi-valued, and $\sigma$ as a function of
$\ddpl$ is shown figure~\ref{fig:2} (left).
The cycle is followed in the clockwise direction and is remarkably similar 
to cycles reported in experiments on lubricants.~\cite{ssglassc}

Let me now comment further on the expression chosen 
for the transformation rates $R_\pm$.
Other prescriptions can be used,~\cite{falk} but introduce some correlations
between free-volume and force fluctuations.
It has been checked that the results presented here do not depend on other possible
choices of $R_\pm$, so long as free-volume activation is incorporated.
The factor $\exp\left[{-{v_0/ v_f}}\right]$ is essential,
and accounts for the fact that at low free-volume, 
activated processes slow down dramatically.
The preceding  expression is chosen because it allows to factorise
$\exp\left[{-{v_0/ v_f}}\right]$ in equations~(\ref{eqn:stz:ddpl}-\ref{eqn:stzdil:3}).
It simplifies somehow the resulting constitutive equations,
but more importantly, the free-volume enters the kinetics of 
shear motion only as much as it determines its time-scale.
The coupling between free-volume relaxation and shear deformation is realized
through this mechanism only, which is a rather minimal and subtle effect.
It is profound and particularly illuminating that such a subtle coupling {\it via}
time-scales suffices to obtain so many non-trivial features.



It is noteworthy that early approach to stick-slip relied
on the introduction of rate-and-state laws, in the spirit of friction laws
used to study earthquakes dynamics.~\cite{carlson96}
In the current approach, free-volume is identified as a logarithmically relaxing
state variable responsible for strain-softening and stick-slip motion.
The importance given to free-volume provides a direct test of the theory
since some appropriate control of the free-volume kinetics could, 
in principle, prevent stick-slip motion.
Although difficult to realize in most experimental set-ups, this might turn out
to have an important practical impact.

The current approach has remained at a qualitative and general level.
This theory has now to pass the test of more stringent
comparison with experimental data.
This might require minor adaptations,
{\it e.g.} to the case of granular materials,~\cite{lemaitre01a,note1}
but the mechanism proposed is of great simplicity and
should adapt to various types of structural systems.
A complete picture of dynamical properties of dense materials will require, however,
to incorporate spatial extension of a sample of material and account for
localization of the deformation into shear bands.
Recent developments in STZ theory have shown a possible mechanism~\cite{langer01}
and should benefit from the incorporation of free-volume dynamics.

This work has benefited from discussions with Jean Carlson, Pascal Favreau,
Delphine Gourdon, Jacob Israelachvili, Jim Langer, and Carl Robert.
I am grateful to Jean Carlson who has permitted my venue in Santa Barbara
and triggered my interest for stick-slip.
I particularly acknowledge Jim Langer as much for his encouragements
as for his stimulating critiques, and for a careful reading of the manuscript.
This work was supported by the W. M. Keck Foundation,
and the NSF Grant No. DMR-9813752,
and EPRI/DoD through the Program on Interactive Complex Networks.


\begin{thebibliography}{10}
\bibitem{ohern01}
C.~S. O'Hern, S.~A. Langer, A.~J. Liu, and S.~R. Nagel, Phys. Rev. Lett. {\bf
  86},  111  (2001).

\bibitem{falk}
M.~L. Falk and J.~S. Langer, Phys. Rev. E {\bf 57},  7192  (1998).
M.~L. Falk and J.~S. Langer, M.R.S. Bulletin {\bf 25},  40  (2000).


\bibitem{lemaitre01a}
{A. Lema\^{\i}tre}, cond-mat/0107422, (2001).

\bibitem{vfgm}
J.~B. Knight {\it et~al.}, Phys. Rev. E {\bf 51},  3957  (1995).
E.~R. Nowak {\it et~al.}, Phys. Rev. E {\bf 57},  1971  (1998).
T. Boutreux and P.~G. de~Gennes, Physica A {\bf 244},  59  (1997).

\bibitem{hassan95}
O.~A. Hassan and M.~C. Boyce, Pol. Engineer. and Science {\bf 35},  331
  (1995).

\bibitem{ssglass}
P.~A. Thompson, G.~S. Grest, and M.~O. Robbins, Phys. Rev. Lett. {\bf 68},
  3448  (1992).
H. Yoshizawa and J. Israechvili, J. Phys.: Condens. Matter {\bf 97},  11300
  (1993).
A.~L. Demirel and S. Granick, Phys. Rev. Lett. {\bf 77},  2261  (1996).

\bibitem{ssglassc}
C. Drummond and J. Israelachvili, Phys. Rev. E {\bf 63},  041506  (2001).

\bibitem{ssgm}
P.~A. Thompson and G.~S. Grest, Phys. Rev. Lett. {\bf 67},  1751  (1991).
S. Nasuno, A. Kudrolli, and J. Gollub, Phys. Rev. Lett. {\bf 79},  949  (1997).
W. Losert, J.-C. G\'eminard, S. Nasuno, and J. P. Gollub, Phys. Rev. E {\bf 61}, 4060 (2000).

\bibitem{note1}
Note that the specificities of a particular material enter {\it via}
the quantities $\bar\mu$ and $R_0$.
In hard-sphere systems, $R_0$ is expected to be proportional to the
collision frequency, $\sqrt{T}/v_f$ (at temperature $T$), 
and the typical normal force, $\bar\mu$, 
is expected to be proportional to the pressure $P$.~\cite{lemaitre01a}
For the sake of simplicity, such specificities are left for future studies, 
and in the present work, $R_0$ and $\bar\mu$ are constants.
This corresponds, for example, to the case when molecules interact {\it via}
a Lennard-Jones potential.

\bibitem{carlson96}
J.~M. Carlson and A.~A. Batista, Phys. Rev. E {\bf 53},  4153  (1996).

\bibitem{langer01}
J.~S. Langer, Phys. Rev. E {\bf 64}, 011504, (2001).

\end{thebibliography}

\end{multicols}
\end{document}